\documentclass[gmd, manuscript]{copernicus}

\begin{document}
\nolinenumbers
\title{PFASST-BISICLES: A Parallel-in-Time Ice Sheet Model}


\Author[1]{Weijia (Emma)}{Liu}
\Author[2]{Hans}{Johansen}
\Author[2]{Daniel F.}{Martin}

\affil[1]{University of Chicago, Chicago, IL, USA}
\affil[2]{Lawrence Berkeley National Laboratory, Berkeley, CA, USA}




\correspondence{Weijia (Emma) Liu (liuwj@uchicago.edu)}

\runningtitle{PFASST-BISICLES}

\runningauthor{Liu, et al}

\received{}
\pubdiscuss{} 
\revised{}
\accepted{}
\published{}


\firstpage{1}

\maketitle

\begin{abstract}
Efficiently simulating ice sheet dynamics on modern computational systems is essential for projecting the response of ice sheets to climate forcing and the resulting contributions to sea level change. On modern computational systems, this demands high levels of parallelism to fully exploit modern architectures. Traditionally, time-dependent simulations like ice sheet models achieve parallelism entirely by distributing the spatial domain, while remaining serial in time, as solutions are marched forward. We present PFASST-BISICLES, which combines the Parallel Full Approximation Scheme in Space and Time (PFASST) with BISICLES ice sheet model to integrate multiple timesteps concurrently. We evaluate numerical agreement and parallel performance using a fully grounded sinusoidal ice stream with adaptive mesh refinement and the MISMIP+ marine ice-sheet benchmark. For the sinusoidal stream, PFASST-BISICLES produces temporal errors approximately an order of magnitude smaller than forward Euler at the same timestep, although both methods exhibit approximately first-order convergence. Spatial scaling levels off near 16 processors, while temporal parallelism provides further acceleration, with nearly linear speedup through approximately 16 temporal processors. In MISMIP+, PFASST-BISICLES reproduces the main thickness and velocity patterns after glacier evolution of 50 years, with the largest differences concentrated near the grounding zone. Increasing the temporal processor count from one to four, with four spatial processors fixed, reduces runtime from 33.8 to 10.5 hours. These results demonstrate that temporal parallelism can complement spatial decomposition in ice-sheet modeling, reducing time to solution while maintaining agreement in the main simulated features.
\end{abstract}


\introduction  

Understanding the dynamic response of the Greenland and Antarctic Ice Sheets to changes in the global climate is essential for projecting impacts like global sea level rise \citep{Edwards-2021}.  Over the past decade or so, a new generation of computationally sophisticated ice sheet models have been developed, which have also been able to take advantage of parallel computing to efficiently use modern computational resources, including the BISICLES model \citep{BISICLES-JCP}.  BISICLES uses block-structured  adaptive mesh refinement (AMR) to improve computational efficiency by dynamically focusing computational resources where needed to sufficiently resolve the relatively fine-scale local dynamics of the grounding lines and ice streams which often control the large-scale dynamics of ice sheets.

At the same time, the trend in high-performance computational architectures has been toward ever-increasing concurrency, often with dozens or even hundreds of cores on a single supercomputer node. \citep{Shalf2020} BISICLES has relied on {\it spatial parallelism} to make use of concurrency -- dividing up the ice sheet spatial domain and distributing the resulting subdomains across processor cores; this works particularly well with block-structured AMR because each refined level is made up of a set of boxes, which are easily distributed. The distributed simulation is then advanced sequentially in time, potentially redistributing the spatial domain as adaptive meshes change to follow the evolving solution. Unfortunately, in part because of the use of AMR to reduce the number of unknowns, BISICLES often exhausts its scalability before it is able to use substantial portions of these resources.  One approach to bringing increased scalability to time-dependent simulations like these is by implementing a {\it parallel in time} scheme \citep{Gander2015, emmett2012toward},  adding some amount of parallelism in the temporal dimension as well.

The concept of adding parallelism to temporal as well as spatial discretizations is not new; the Parareal method was first proposed by Lions in 2001 \citep{Lions2001}. Another approach, Multigrid Reduction in Time (MGRIT) is based on concepts drawn from multigrid solvers commonly used to solve linear systems. \citep{FalgoutEtAl2014_MGRIT} In this work, our approach is based on the Parallel Full Approximation Scheme in Space and Time (PFASST) approach presented in \cite{emmett2012toward}.

\section{Method}
Our approach is to begin with the existing BISICLES ice sheet model \citep{BISICLES-JCP}, retaining its spatial discretizations, but replacing the original time discretization with PFASST. This entailed some reorganization of the code to add the ability to compute the right-hand-side of the thickness update equation directly, in contrast to the integrated spatio-temporal scheme described in 
\cite{BISICLES-JCP}.

\subsection{Spatial Discretization and Adaptivity}

As described in \cite{BISICLES-JCP}, BISICLES uses a modified form of the vertically-integrated formulation of the momentum balance presented in \cite{Schoof10}. 
\begin{gather}
\dfrac{\partial h}{\partial t}+\nabla\cdot(\boldsymbol{u}h) = M_s+M_b,\label{eq:dhdt} \\ 
\beta^2(u_b) \vec{u_b} + \vec{\nabla} \cdot [h \mu(\dot \epsilon^2,T)(\vec{\nabla} + \vec{\nabla}^T) \vec{u_b} - 2 \mu(\dot \epsilon^2,T)(\vec{\nabla} \cdot \vec{u_b})] = -\rho g h \vec{\nabla}s, \label{eq:momentum}
\end{gather}
where $h$ is ice thickness, $\boldsymbol{u}(x,y)$ is the vertically-integrated velocity vector, $\vec{u_b}$ is the ice velocity at the base of the ice, $M_s$ and $M_b$ are surface accumulation rate and basal melt rate, $\beta^2$ is the coefficient of friction at the base of the ice (which is 0 under floating ice shelves), $s$ is the upper surface of the ice, $\rho$ is the ice density, $g$ is the gravitational acceleration (9.81 $\frac{m}{s^2}$), and $\mu$ is ice viscosity, which is a function of the strain rate tensor $\dot{\boldsymbol{\epsilon}}$ and the ice temperature $T$.

\subsection{Sequential Time Discretization}
The time integration scheme described in \cite{BISICLES-JCP} is a predictor-corrector scheme in cell-centered solution values at time $t^n$ are used to predict face-centered values for the ice thickness at the half time ($t^{n+\frac{1}{2}} = t^n + \frac{1}{2} \Delta t$ using an unsplit scheme based on the Piecewise Parabolic Method (PPM) \citep{PPM}. These face-centered thicknesses are then multiplied by face-centered velocities to compute ice fluxes at the half-time, which are then used to compute a finite-volume divergence of the fluxes. Although the use of fluxes centered at the half time would define a scheme which is second-order in time, the use of velocities computed at $t^n$ results in a scheme which is formally first-order accurate in time, although it does maintain conservation.

\subsection{PFASST Parallel-in-Time Discretization}

The PFASST algorithm was first introduced by \cite{emmett2012toward} to solve generic initial value problems for Partial Differential Equations (PDEs). In our model, we apply this algorithm to a two-dimensional mass transport equation in the Picard integral form
\begin{equation}
    h(t)=h_0+\int_0^{t}f(h(\tau),\vec{u}(\tau))d\tau, t\in(0,T_f], \label{eq:dhdt_rhsf}
\end{equation}
where $h_0$ is the initial ice thickness, the right-hand side function $f(h,\vec{u})$ and the velocity field $\vec{u}$ are obtained from equation \ref{eq:dhdt} and equation \ref{eq:momentum}, respectively.



The fundamental concept underlying PFASST involves the hierarchical discretization of the temporal dimension, creating a structured hierarchy of temporal grids commonly referred to as "temporal levels", varying from coarse to fine. Within each of these levels, we employ the Spectral Deferred Correction (SDC) technique \citep{dutt2000spectral,minion2003semi} to compute the solutions within each physical step by iteratively approximating the integrals in equation \ref{eq:dhdt_rhsf} on quadrature nodes using lower-order methods. The effective inter-level communication is achieved by using the Full Approximation Scheme (FAS) \citep{brandt1977multi,trottenberg2000multigrid}, which is discussed at the end of this section.

For a simulation time span $[0,T_f]$, the temporal direction is first uniformly discretized into a grid consisting of $N$ intervals. Subsequently, we assign different temporal levels to each time interval $[t_n,t_{n+1}]$, where the coarsest and finest levels are designated as level 0 and level $L$, respectively. On each level $l\in[0,L]$, $[t_n,t_{n+1}]$ is further divided into $M_l$ sub-intervals with $(M_l+1)$ SDC nodes. Hence, for any level $l$, the division of time steps can be summarized as
\begin{align}
0&=t_{0}<t_{1}<...<t_{N}=T_f, \Delta t=t_{n+1}-t_n,\\
t_n&=t_{n0}<t_{n1}<...<t_{nM_l}=t_{n+1}, \Delta t_m=t_{n(m+1)}-t_{nm}.
\end{align}

Within one substep $[t_{nm},t_{n(m+1)}]$, we use explicit Euler to compute solutions $h$ on each level iteratively. For notational simplicity, we denote the solution $h$ evaluated at SDC node $t_{nm}$ at SDC iteration $k$ as $h_{nm}^k$, likewise for velocity $u_m$ and right-hand size function $f(h_{nm}^k,u_{nm})$. The recursive equation to compute $[h^{k+1}_1,...,h_{M_l}^{k+1}]$ at SDC nodes $[t_{n1},...,t_{nM_l}]$ can be then written as
\begin{equation}
    h_{n(m+1)}^{k+1}=h_{nm}^{k}+\Delta t_m\left(f(h_{nm}^{k+1},u_{nm})-f(h_m^k,u_m)\right)+I_{m}^{m+1}(h^k,\vec{u}),\ \ m=0,...,M_l-1, \label{eq:sdc_sweep}
\end{equation}
where $I_m^{m+1}(h^k,\vec{u})$ is the numerical quadrature approximation to the right-hand side function integral in $[t_{nm},t_{n(m+1)}]$ at $k$th iteration
\begin{equation}
    \int_{t_{nm}}^{t_{n(m+1)}}f(h^k,\vec{u})d\tau\approx I_m^{m+1}(h^k,\vec{u})=\Delta t\sum_{j=0}^{M} q_{mj}f(h_{nj}^k,u_{nj}).
\end{equation} 
The coefficients denoted as $q_{mj}$ are the quadrature weights at SDC nodes and can be pre-computed. Thus, the above quadrature approximation is a simple matrix-vector multiplication. The compact form of equation \ref{eq:sdc_sweep} is then 
\begin{equation}
    \vec{h}^{k+1}=\vec{I}h_0+\vec{S}(\vec{F}^{k+1}-\vec{F}^k)+\Delta t\vec{Q}\vec{F}^k, \label{eq:collocation_SDC}
\end{equation}
where $\vec{I}$ is the identify matrix, $\vec{Q}$ is the quadrature integral matrix of size $M_l\times(M_l+1)$ with entries $q_{mj}$, $\vec{h}$ a vector of ice thickness evaluated at all SDC nodes $[h_{n0},...,h_{nM_l}]^T$, $\vec{F}$ a vector of the right-hand side function evaluated at all SDC nodes $[f(h_{n0},u_{n0}),...,f(h_{nM_l},u_{nM_l})]^T$, and $\vec{S}$ is defined as
\begin{align}
    s_{ij}=
    \begin{cases}
     \Delta t_{(j-1)}&\text{if } j\leq i,\\
     0 &\text{otherwise}.
    \end{cases}
\end{align}

Typically, SDC nodes are selected to align with standard Gaussian quadrature nodes, such as Gauss-Legendre nodes and Gauss-Radau nodes. Here, we opt for Gauss-Lobatto nodes on the finest level. The coarse SDC nodes are then chosen to be a subset of the fine SDC nodes to allow easier inter-level communications. It is worth noting, however, that this choice does not guarantee that the SDC nodes at the coarser level will also be Gauss-Lobatto nodes, resulting in a decrease in the order of accuracy \citep{emmett2012toward}. In this study, we choose 2 and 3 nodes for the coarse and fine level. 

In PFASST, SDC iterations are performed on multiple levels with a focus on conducting additional iterations on coarse levels to reduce the total work load. This approach is also called multi-level SDC \citep{speck2015multi}. While this approach effectively curtails computational costs, it inevitably introduces a trade-off with respect to solution accuracy. 
Hence, to obtain the precision achieved by fine SDC iterations while operating at the coarser level resolution and also to transfer solutions between levels, an FAS correction term $\tau$ is introduced to correct the solutions on the coarse level
\begin{equation}
    \vec{\tau}=\Delta t(\vec{Q}^C\vec{F}^C-\vec{T}_F^C\vec{Q}^F\vec{F}^F),
\end{equation}
where the superscripts $C$ and $F$ denote coarse and fine, respectively, and $\vec{T}_F^C$ is the restriction operator that transfers the quantities from fine level to coarse level. Because we choose the SDC nodes on the coarse level to be a subset of the fine level nodes, the restriction operator in our case is simply a pointwise injection. The collocation problem (equation \ref{eq:collocation_SDC}) within the time interval $[t_n,t_{n+1}]$ with FAS correction on coarse level can be then expressed as
\begin{equation}
    \vec{h}^{k+1}=\vec{I}h_0+\vec{S}(\vec{F}^{k+1}-\vec{F}^k)+\Delta t\vec{Q}\vec{F}^k+\vec {\tau}^k. \label{eq:collocation_SDC_FAS}
\end{equation}

Upon the completion of the coarse level iterations, the solutions are transferred from coarse level to fine level using an interpolation operator $\vec{T}_C^F$. In this context, $\vec{T}_C^F$ is constructed using standard polynomial interpolation.

\subsection{Spatial and Temporal Parallelization}
In our model, we use two MPI-based communicators for spatial parallelization on block-structured AMR grids and temporal parallelizations. Thus, every processor is a member of both communicators, i.e., it holds a sub-box of the AMR grids in BISICLES at a specific time step in PFASST. 

\begin{figure}[h]
\includegraphics[width=0.7\linewidth]{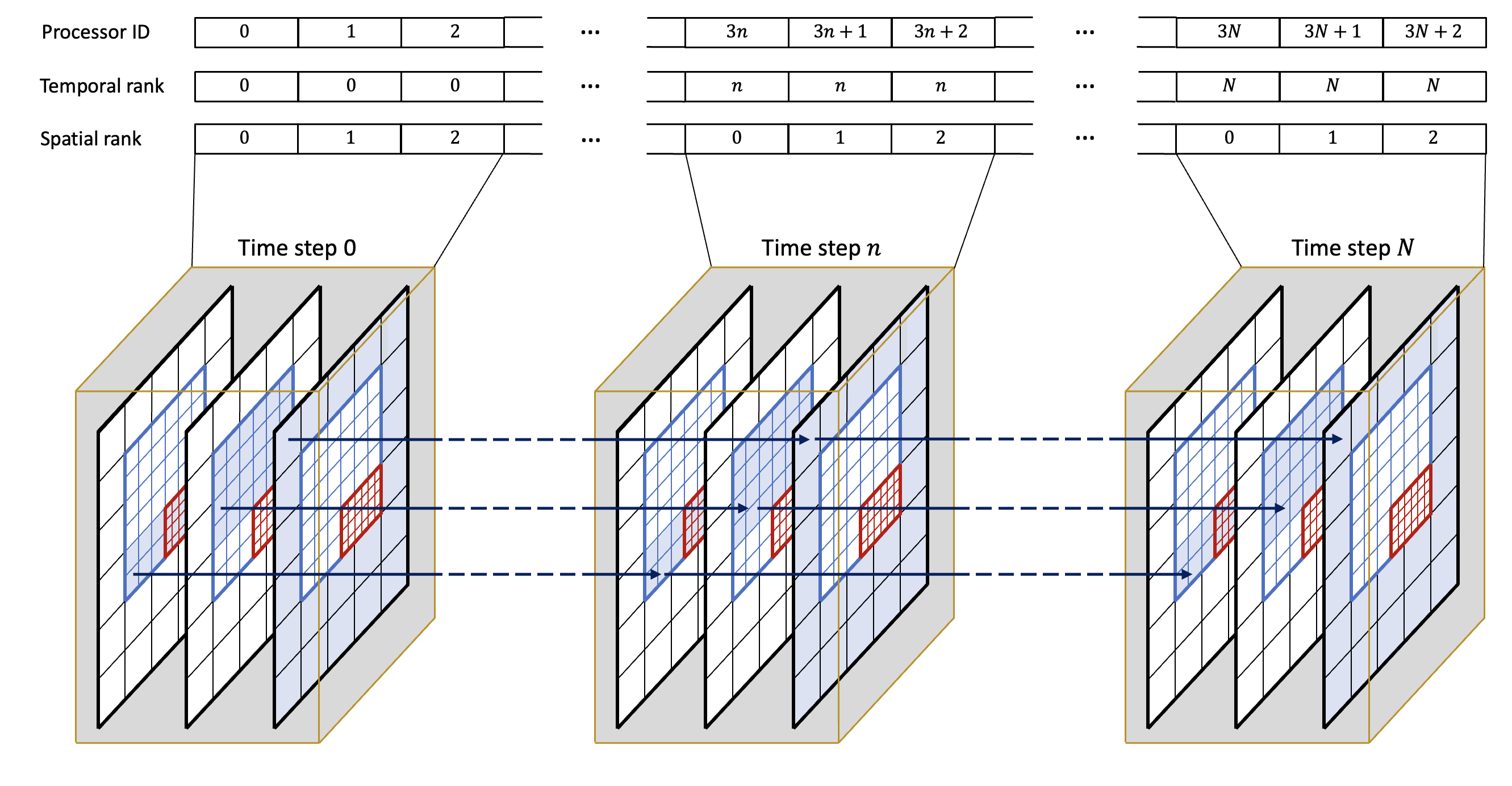}
\caption{Example of parallelization with $N$ temporal processors with each temporal processor contains 3 spatial processors. The temporal direction is decomposed into $N$ time steps using a PFASST communicator, where each time step is represented as one gray box. Within each time step, the spatial grid is decomposed into 3 sub-boxes and managed by 3 spatial processors, depicted as 3 slices. Each slice is responsible for one sub-box of the entire grid,  represented in shaded blue areas. Collectively, these slices collectively compose the entirety of the spatial domain. The communication across different time steps is achieved by slice-to-slice packing/unpacking as depicted in the blue arrows. In this example, the spatial grid has 3 levels of AMR from coarse to fine shown as black, blue, and red grids.}
\label{fig:MPI_distribution}
\end{figure}

In the diagram presented in Figure \ref{fig:MPI_distribution}, we employ a rigorous decomposition strategy for the temporal and spatial direction. First, the temporal direction is decomposed onto $N$ MPI-temporal-ranks using a PFASST communicator that represents physical time steps, shown as the gray boxes. Then, within each of these time steps, the entire spatial grid is encapsulated and subsequently distributed across 3 MPI-spatial-ranks, coordinated by a local BISICLES communicator, represented as 3 slices within each gray box. The data transfer between temporal processors is confined exclusively to those residing within the same spatial rank, realized through slice-to-slice transfers as depicted by the blue arrows. In this example, we show a spatial grid division into 3 sub-boxes (shaded blue areas) with 3 AMR levels (black, blue, and red grids). This method for distributing temporal and spatial processors can be generalized to any number of boxes and AMR levels. In essence, this approach implemements the concurrent operation of "$N$ ice sheets" in parallel, where $N$ is the total number of physical time steps. 

To implement the PFASST algorithm into the BISICLES ice sheet model, we used the PFASST library and constructed a BISICLES vector factory and binding interfaces to ensure data compatibility. 

\section{Results}
We test our implementation on two problems: a sinusoidal ice stream \citep{BISICLES-JCP}, and the MISMIP+ marine ice sheet community benchmark \citep{MISMIP+-design}. 
These test cases serve as complementary benchmarks for assessing the capabilities of our implementation. The sinusoidal ice stream problem serves as a foundational test, primarily employed to showcase the efficacy of our parallel-in-time ice sheet model. MISMIP+ focuses on grounding line retreat and advance and is useful to assess the suitability of the PFASST-BISICLES model for a rapidly changing ice sheet with moving grids and a solution regime which has been demonstrated to be sensitive to under-resolution. 

To compare the difference between the traditional sequential model and parallel-in-time model, we conducted two sets of simulations for each of the aforementioned test problems. Henceforth, we will refer to these as the "sequential run" and the "parallel run". 

\subsection{Sinusoidal ice stream}
Following \cite{BISICLES-JCP}, we examine a sinusoidal ice stream on an inclined 
plane, with the bed tilted at -0.5 degree in the $x$ direction. The domain is set to be $[160,160]$ km with periodic boundary conditions on all sides. The initial ice thickness is set as 1000 m across the domain. The ice is fully grounded 
and  a linear basal friction law is applied.

\begin{figure}[h!]
\includegraphics[width=1.0\linewidth]{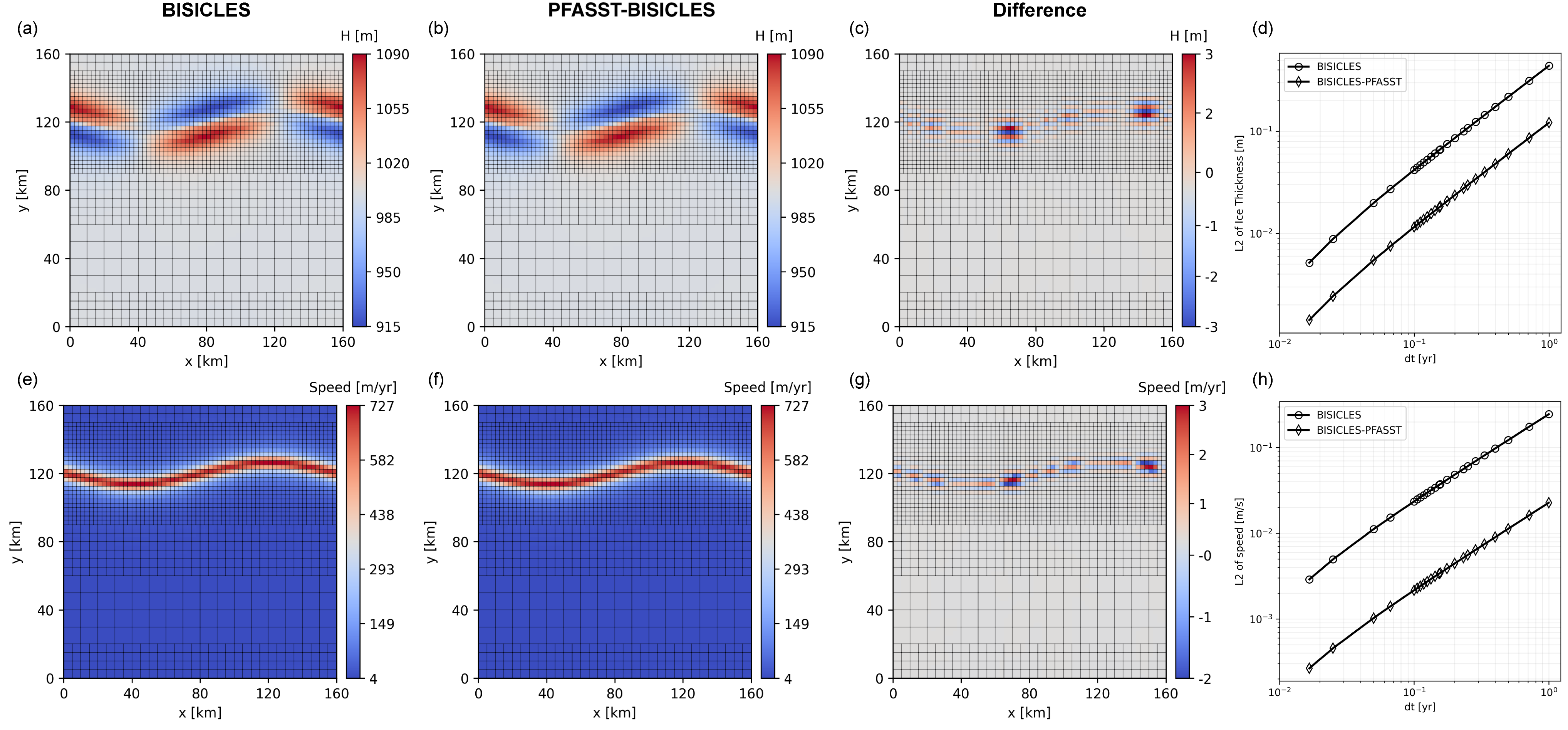}
\caption{Comparison of sinusoidal ice stream problem using both BISICLES and PFASST-BISICLES algorithms. The first and second rows show the results and difference of ice thickness and surface speed, respectively. The (a), (b), and (c) columns show the results using BISICLES, PFASST-BISICLES, and their differences. Panel (d) shows the temporal convergence using $dt=0.05$ year as a reference. All simulations use three levels of AMR grids and a total simulation time of 20 years.}
\label{fig:twisty_stream_results}
\end{figure}

For this problem, we utilize three levels of AMR grids in both simulations, with the coarsest AMR level (level 0) having a grid size of $\Delta x=16$ km and the finest level (level 2) having a grid size of $\Delta x=4$ km. We use the Jacobian-free Newton-Krylov method to solve the force balance equation \ref{eq:momentum} \citep{brown1990hybrid,knoll2004jacobian}. For the temporal scheme, we use both forward Euler (BISICLES) and PFASST algorithm (PFASST-BISICLES) to solve the thickness evolution equation \ref{eq:dhdt}.

Figure \ref{fig:twisty_stream_results} presents a comparison of the BISICLES and PFASST-BISICLES 
simulations after 20 years. We compare the thickness (first row) and surface speed (second row). The results from BISICLES and PFASST-BISICLES are shown in the first and second columns, respectively, with the differences highlighted in the third column. The maximum difference observed is approximately 0.28\% for ice thickness and 0.41\% for surface speed. The fourth column illustrates the convergence of both methods for ice thickness and surface speed, using $dt=0.05$ year as reference. The L2 error for PFASST-BISICLES is about an order of magnitude lower than that of BISICLES at the same temporal resolution due to the SDC correction scheme, though both methods exhibit a convergence rate with a slope of approximately 1.

\begin{figure}[h!]
\includegraphics[width=0.9\linewidth]{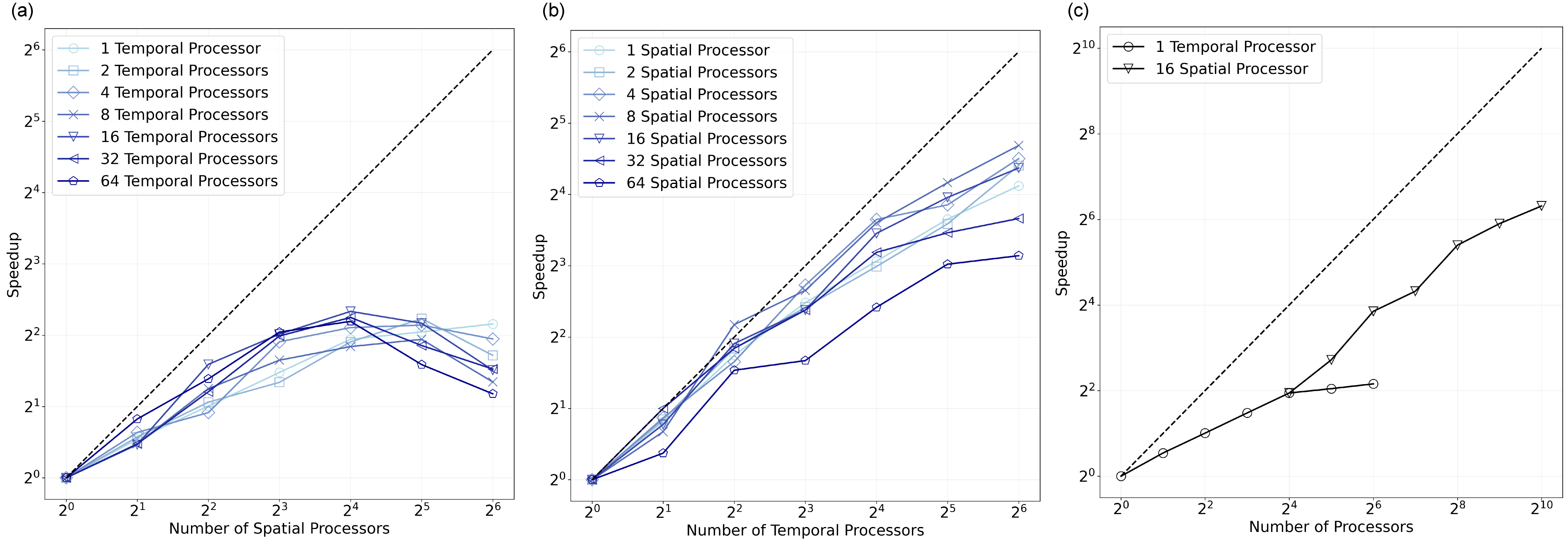}
\caption{Scalability of the PFASST-BISICLES algorithm for sinusoidal ice stream simulations. The figures (a) and (b) show the spatial and temporal scalability. The figure (c) illustrates how adding temporal parallelization enhances  overall scalability compared to relying solely on spatial parallelization.}
\label{fig:twisty_stream_scalability}
\end{figure}

We then compare the scalability of BISICLES and PFASST-BISICLES on the same sinusoidal ice stream problem. The original BISICLES implementation 
allows parallelization only in the spatial dimension, while PFASST-BISICLES enables parallelization in both spatial and temporal dimensions. We conducted simulations using up to 64 processors in both the spatial and temporal dimensions, recording the computational time for each simulation to complete. The computational times for simulations using more than one processor in a given parallelization direction were normalized against the computational time of a single processor in that direction. The resulting speedup plots are presented in Figure \ref{fig:twisty_stream_scalability}, with solid lines representing the normalized speedup and dashed lines representing ideal linear speedup.

Figure \ref{fig:twisty_stream_scalability} (a) illustrates the spatial scalability of PFASST-BISICLES. The computational time decreases significantly as more processors are added, but after about 16 processors, the decrease in computational time slows and even begins to slightly increase. This indicates that PFASST-BISICLES has reached its maximum efficiency in spatial parallelization for this particular problem, meaning that adding additional processors beyond this point does not further reduce computational time.

However, further reductions in computational time can be achieved by leveraging temporal parallelization, as shown in Figures \ref{fig:twisty_stream_scalability} (b) and (c). The computational time decreases almost linearly with the number of processors added to the temporal dimension until the number of temporal processors reaches 16. Beyond this point, adding more processors continues to reduce computational time, though at a diminishing rate. This trend holds across all levels of spatial parallelization, demonstrating that PFASST-BISICLES exhibits strong scalability in the temporal dimension, while combining spatial and temporal parallelism leads to significant improvements in computational efficiency and time-to-solution.

\subsection{MISMIP+}

To assess PFASST-BISICLES in a more dynamically challenging setting, we consider MISMIP+ \citep{MISMIP+-design}, a community benchmark for marine ice sheet response to subshelf melting. Its idealized geometry represents an ice stream feeding a laterally confined ice shelf, with qualitative similarities to Pine Island Glacier in West Antarctica. Changes in melting modify shelf buttressing and drive grounding-line migration, testing the coupled evolution of ice thickness and velocity in response to temporally dynamic forcing. Here, we compare BISICLES and PFASST-BISICLES solutions after 50 years on a uniform 1 km grid covering an $800 \times 80$~km domain. While the full MISMIP+ protocol extends the melting phase to 100 years and then allows for recovery, we chose to examine solutions after 50 years because that is in the middle of the dynamically active regime, presenting a more challenging test in terms of temporal accuracy.

\begin{figure}[h]
\includegraphics[width=1\linewidth]{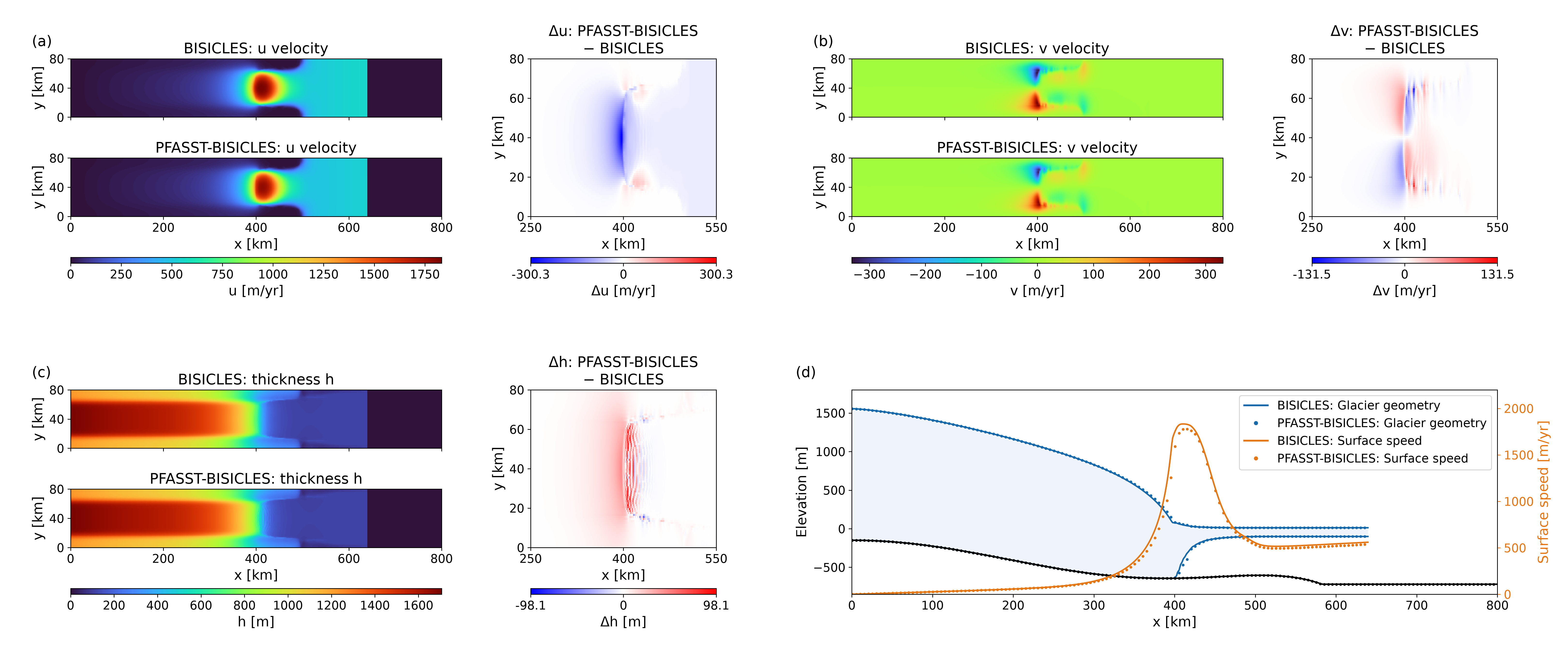}
\caption{BISICLES and PFASST-BISICLES solutions for MISMIP+ at 50 years with 1 km grid spacing. Panels (a)-(c) show along-flow velocity $u$, transverse velocity $v$, and thickness $h$, respectively. Each panel includes both full-domain solutions and their difference (PFASST-BISICLES $-$ BISICLES) at $x\in[250,550]$ km. Panel (d) shows ice geometry and surface speed along the domain centerline ($y=40$ km); solid lines denote BISICLES and dots denote PFASST-BISICLES.}
\label{fig:mismip_results}
\end{figure}

Figure~\ref{fig:mismip_results} shows that PFASST-BISICLES reproduces the main flow pattern of the MISMIP+ problem, such as rapid thinning of ice thickness near the grounding line at x$\sim 400$ km, concentrated fast flow in the center, and transition from grounded ice to floating shelf as the grounding line retreats. Taking BISICLES as the reference, the domain-wide relative $L_2$ differences are approximately $0.7\%$ for thickness $h$ and $6.1\%$ for the horizontal velocity $u$. The largest discrepancies occur near the grounding zone around $x=400$ km, where thickness and velocity vary rapidly. The largest difference between BISICLES and PFASST-BISICLES are approximately 98 m in thickness and 300 m/yr in $u$ velocity. The centerline profiles show similar ice geometry (panel d), while PFASST-BISICLES produces a slightly lower speed maximum downstream. 

\begin{figure}[h]
\includegraphics[width=0.6\linewidth]{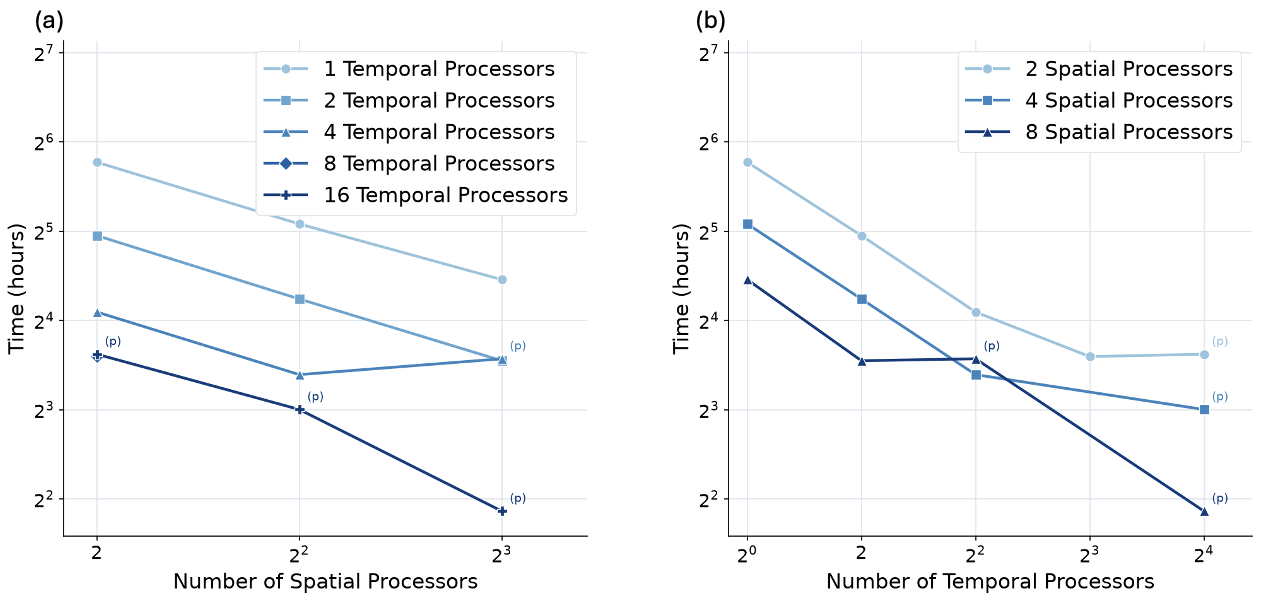}
\caption{Wall-clock runtime of BISICLES and PFASST-BISICLES for MISMIP+ simulations. (a) spatial scaling at fixed temporal processor counts. (b) temporal scaling at fixed spatial processor counts. Points marked ``(p)'' were obtained on a different machine.}
\label{fig:mismip_scalability}
\end{figure}

Figure~\ref{fig:mismip_scalability} shows computational performance by varying the spatial and temporal processor counts. With one temporal processor, increasing the number of spatial processors from two to eight reduces runtime from 54.7 to 22.0 h, giving a speedup of 2.5. Temporal parallelization provides additional runtime reductions. With four spatial processors, increasing the temporal processor count from one to four reduces runtime from 33.8 to 10.5 h, corresponding to a speedup of 3.2. With two spatial processors, increasing the temporal allocation from one to eight reduces runtime from 54.7 to 12.1 h, a speedup of 4.5. These comparisons demonstrate that parallelizing the time dimension can substantially reduce time to solution for this problem, although the gains are sublinear. The shortest recorded runtime is 3.6 h with eight spatial and sixteen temporal processors. This configuration, together with the other points marked ``(p)'', was run on a different machine. Comparisons involving these points therefore include hardware effects and cannot be interpreted solely as gains from increased parallelization.

\subsection{Discussion}
The two test cases both provide computational accuracy and efficiency assessments from incorporating temporal parallism in PFASST-BISICLES. The fully grounded sinusoidal stream, from \citet{BISICLES-JCP}, tests temporal integration on an AMR hierarchy without grounding-line migration. PFASST-BISICLES produces errors approximately an order of magnitude below forward Euler at the same timestep, while both methods exhibit approximately first-order convergence. Although PFASST supports high-order integration \citep{emmett2012toward}, these results establish a reduction in error magnitude even with formally first-order integration in the tested regime. The observed convergence rate motivates further examination of timestep size, iteration tolerances, and the coupling between thickness evolution and velocity updates.

MISMIP+ extends to a marine ice sheet with strong spatial gradients and grounding line dynamics. Using PFASST-BISICLES, we can reproduce the overall glacier geometry and flow velocity after 50 year of dynamic thinning compared to the baseline BISICLES implementation. While the agreement is weaker near the grounding zone, it is likely that this is indicative of the improved temporal accuracy of the PFASST-BISICLES scheme demonstrated in the previous example. 
The results therefore support the reproduction of the large-scale flow pattern and dynamic response, while indicating that further study of a comparison of grounding-line trajectories is needed to assess comparative accuracy of the retreat history.

The scalability results identify a practical benefit shared by both problems. For the sinusoidal stream, spatial scaling levels off near 16 processors, while temporal parallelization provides further acceleration, with nearly linear speedup through approximately 16 temporal processors. This behavior is consistent with earlier PFASST applications that extended spatial strong scaling for particle simulations and the three-dimensional heat equation \citep{speck2012massively,bader2014space}. The runtime reductions in MISMIP+ show that temporal parallelization remains useful in a more dynamically demanding setting, although efficiency gains are somewhat weaker than the simple steady state sinusoidal stream case.

The performance results of MISMIP+ case also distinguish shorter runtime from lower computational cost. With four spatial processors, increasing the temporal allocation from one to four gives a speedup of approximately 3.2 while using four times as many processors. This corresponds to roughly 25\% more processor-hours. While the faster simulation carries an additional resource cost, this will often represent a more-efficient use of available computational resources on modern multicore architectures. 

\conclusions  
In this paper, we coupled the PFASST parallel-in-time algorithm with the BISICLES ice sheet model and evaluated the resulting method on two ice-flow examples. The sinusoidal stream demonstrates compatibility of PFASST-BISICLES with an AMR hierarchy.
PFASST-BISICLES consistently shows lower error than the original BISICLES model under the same temporal resolution, even though formal convergence is still first-order in time. Using the MISMIP+ benchmark, we extended the assessment to marine ice-sheet dynamics, where PFASST-BISICLES reproduces the large-scale dynamic response in ice geometry and flow, despite somewhat weaker agreement near the grounding zone.

These experiments both demonstrate the practical value of adding temporal parallelism to improve computational efficiency. For the sinusoidal stream, distributing work across timesteps allows reduce runtime with larger numbers of processor cores after spatial scaling is exhausted. The MISMIP+ experiments also show substantial runtime reductions, including a speedup of 3.2 with four temporal processors at a fixed spatial allocation. Temporal parallelism therefore provides an additional means of accelerating both fully grounded and marine ice-sheet simulations.

This PFASST-BISICLES example demonstrates how incorporating temporal parallelism can expand the efficient use of parallel computing in ice-sheet modeling by distributing work across both space and time. This capability is relevant to ice-sheet projections, where exploring uncertainty in basal conditions and climate forcing requires repeated simulations, each potentially expensive and time-consuming due to sequential time integration over long solution times. By reducing the turnaround time of individual runs, the approach can support broader parameter studies and more detailed simulations, connecting advances in parallel time integration to the computational demands of ice-sheet prediction.




\codedataavailability{Code and examples for this work are available via the NERSC portal at {\url{https://portal.nersc.gov/cfs/m1411/pfasst-bisicles}}} 





\noappendix       







\authorcontribution{WL performed the coding, ran the examples, and wrote most of the text. HSJ provided advice regarding parallel in time and PFASST schemes. DFM assisted with coding, text editing, and provided support.} 

\competinginterests{The authors declare that they have no competing interests.} 


\begin{acknowledgements}
Support for this work was provided through the Scientific Discovery through Advanced Computing (SciDAC) program funded by the U.S. Department of Energy (DOE), Office of Science, Biological and Environmental Research and Advanced Scientific Computing Research programs, as a part of the ProSPect SciDAC Partnership. Work at Berkeley Lab was supported by the Director, Office of Science, of the U.S. Department of Energy under Contract No. DE-AC02-05CH11231. Part of WL's work was support by Schmidt AI in Science Fellowship. This research used resources of the National Energy Research Scientific Computing Center (NERSC), a U.S. Department of Energy Office of Science User Facility located at Lawrence Berkeley National Laboratory, operated under Contract No. DE-AC02-05CH11231 using NERSC award ASCR-ERCAPm1041.
\end{acknowledgements}







\bibliographystyle{copernicus}
\bibliography{references.bib}

\end{document}